\documentclass[aps,asmsymb,twocolumn,showpacs,amsmath,prb]{revtex4-1}

\usepackage{graphicx}
\usepackage{subfigure}
\usepackage{epsfig}
\usepackage{dcolumn}
\usepackage{bm}
\usepackage{color}
\usepackage{xcolor}
\usepackage{bm}
\usepackage{hyperref}
\usepackage{amsmath, amssymb, dsfont}

\def\be{\begin{equation}}       \def\ee{\end{equation}}
\def\bea{\begin{eqnarray}}      \def\eea{\end{eqnarray}}
\def\ba{\begin{array}}
\def\ea{\end{array}}
\def\bnum{\begin{enumerate} }
\def\enum{\end{enumerate}}

\def\=>{\Rightarrow}
\def\>{\rightarrow}

\def\eye2{Fathbb{I}}
\def\bk{{\bf k}}

\def\bJ{{\bf J}}
\def\bL{{\bf L}}
\def\bn{{\bf n}}
\def\bS{{\bf S}}
\def\br{{\bf r}}
\def\bv{{\bf v}}
\def\si{{\sigma}}

\def\bS{{\bf S}}

\renewcommand{\v}[1]{{\bf #1}}

\renewcommand{\>}{\rangle}

\newcommand{\eps}{\epsilon}

\newcommand{\al}{\alpha}

\begin{document}

\title{Imbert-Fedorov shift in pseudospin-$N/2$ semimetals and nodal-line semimetals}
\author{Yi-Ru Hao}
\author{Luyang Wang}
\email{wangluyang730@gmail.com}
\author{Dao-Xin Yao}
\email{yaodaox@mail.sysu.edu.cn}
\affiliation{State Key Laboratory of Optoelectronic Materials and Technologies, School of Physics, Sun Yat-Sen University, Guangzhou 510275, China}

\begin{abstract}
The Imbert-Fedorov (IF) shift is the transverse shift of a beam at a surface or an interface. It is a manifestation of the three-component Berry curvature in three dimensions, and has been studied in optical systems and Weyl semimetals. Here we investigate the IF shift in two types of topological systems, topological semimetals with pseudospin-$N/2$ for an arbitrary integer $N$, and nodal-line semimetals (NLSMs). For the former, we find the IF shift depends on the components of the pseudospin, with the sign depending on the chirality. We term this phenomenon the pseudospin Hall effect of topological fermions. The shift can also be interpreted as a consequence of the conservation of the total angular momentum. For the latter, if the NLSM has both time-reversal and inversion symmetries, the IF shift is zero; otherwise it could be finite. We take the NLSM with a vortex ring, which breaks both symmetries, as an example, and show that the IF shift can be used to detect topological Lifshitz transitions. Finally, we propose experimental designs to detect the IF shift.
\end{abstract}
\date{\today}
\maketitle	

\section{Introduction}
When a beam of light is reflected at a surface or interface, it experiences a chirality-dependent transverse shift\cite{Imbert1972,Fedorov1955}. This phenomenon, named the Imbert-Fedorov (IF) shift after its discoverers, originates from the chirality-dependent Berry curvature of photons that can be derived from Maxwell equations\cite{Onoda2004,Bliokh2006}. Semiclassically, a wave packet of photons with certain chirality gains anomalous velocity during the reflection process, which is the cross product of the Berry curvature and an effective force acting on it, resulting in the IF shift. Chirality is also a conserved quantity for Weyl fermions, the Berry curvature of which is also chirality-dependent. Therefore, the IF shift can also occur in the recently discovered Weyl semimetals (WSMs)\cite{Armitage2018} whose low energy quasiparticles are Weyl fermions\cite{Jiang2015,Yang2015,Wang2017}. In this work, we generalize the investigation of the IF shift to other types of topological semimetals, namely, the pseudospin-$N/2$ semimetals with $N$ being an arbitrary positive number, and nodal-line semimetals (NLSMs).

Pseudospin refers to the quantum degree of freedom (DOF) that quasiparticles possess in addition to their orbital DOF, and behaves in a manner that is mathematically equivalent to spin\cite{Pesin2012}. For instance, the sublattice DOF in graphene can be viewed as a pseudospin\cite{CastroNeto2009}. In three dimensions, the WSMs have the effective Hamiltonian $H_w=\bk\cdot{\bf\sigma}$, where the Pauli matrices ${\bf\sigma}$ may also denote a degree of freedom which is not related to spin, and hence can be viewed as a pseudospin. As the spin DOF is used in spintronics\cite{Zutic2004,Pesin2012}, the pseudospin DOF finds its applications in pseudospintronics\cite{Pesin2012}.

Although pseudospin behaves like spin, there can be distinct properties associated with it, one of which is that it does not obey spin-statistics theorem. In particle physics, particles obey the Poincar\'e group symmetry, from which spin-statistics theorem is derived. It tells us that fermions possess half-integer spin while bosons possess integer spin. In condensed matter physics, on the other hand, crystalline solids respect symmetries of the space groups which are subgroups of the Poincar\'e group and are less constrained, and the pseudospin DOF is not constrained by spin-statistics theorem\cite{Bradlyn2016}. As a result, fermions can possess integer pseudospin while bosons can possess half-integer pseudospin. Examples for the former include pseudospin-1 fermions\cite{Bradlyn2016,Ezawa2016,Tang2017,Zhu2017}, and that for the latter include Weyl magnons\cite{Li2016,Mook2016,Su2017,Owerre2018,Jian2018}, photonic\cite{Lu2013,Lu2015,Wang2016,Yang2018} and phononic crystals with Weyl points\cite{Xiao2015,Rocklin2016,Zhang2018}, and so on. It has been shown topological fermions with pseudospin-$N/2$ with $N=1,2,3$ can be protected by space group symmetries in three dimensions\cite{Bradlyn2016}. In this work, we adopt the semiclassical equations of motion (EOMs) to derive the IF shift of the topological fermions with an arbitrary pseudospin. Then we interpret the IF shift as a consequence of the conservation of the total angular momentum, which is the sum of the orbital angular momentum and the pseudospin angular momentum. Our results show that the IF shift is related to the pseudospin component of the topological fermions, thus we term the phenomenon the pseudospin Hall effect (PSHE).

In the pseudospin-$N/2$ semimetals, the quasiparticles' momentum couple linearly with the pseudospin, giving rise to isotropic linear dispersion. We go beyond such semimetals next, to NLSMs, which host band crossings along a line or a ring in the Brillouin zone, and are highly anisotropic. NLSMs are protected by spatial symmetries, along with time-reversal symmetry in some cases, and can be classified according to their symmetries\cite{Fang2016}. NLSMs with different symmetries have different low energy effective Hamiltonian, hence different forms of Berry curvature and IF shift. We focus on two types of NLSMs, one with both time-reversal and inversion symmetries, and another with neither of the two symmetries, and investigate the IF shift in these two systems. In the former, the IF shift vanishes; while in the latter, we find that the IF shift is finite and can be used to detect topological Lifshitz transitions.

Finally, we design an experiment to detect the IF shift. While we discuss the IF shift in the context of topological fermionic systems, the calculations and results can also be applied directly to bosonic systems with arbitrary pseudospins and with nodal lines. As such, when addressing the detection of this effect, we design the experimental setup in both an electronic system and a photonic system.

Our paper is organized as follows. In Sec. \ref{SecII}, we discuss the pseudospin Hall effect in pseudospin-$N/2$ semimetals. In Sec. \ref{SecIII}, we calculate the IF shift in two types of NLSMs, and discuss its application for the detection of topological Lifshitz transitions. We propose an experimental design in Sec. \ref{SecIV}. Finally, in Sec. \ref{SecV}, we give a summary.

\section{Pseudospin Hall effect in pseudospin-$N/2$ semimetals}\label{SecII}
\subsection{Pseudospin-orbit coupled Hamiltonian}
We study the topological fermions whose Hamiltonian can be written as $H=v\bk\cdot\bS$ where $\bS$ is the spin-$N/2$ representation of the SO(3) group, satisfying the angular momentum algebra $[S_i,S_j]=i\eps_{ijk}S_k$. Here, $\bS$ represents the pseudospin degree of freedom, thus $N$ can be either odd or even. The eigenvalues and eigenstates can be found in the following way. We write $H/(vk)=\hat{\bk}\cdot {\bf S}$ where $\hat{\bf k}=(\sin\theta\cos\varphi,\sin\theta\sin\varphi,\cos\theta)$ is the direction of $\bk$ in the spherical coordinate system. We can diagonalize $\hat{\bk}\cdot {\bf S}$ by a unitary transformation\cite{He2012}, with the unitary matrix $P$. We use the convention that $S_z$ is a diagonal matrix, i.e. $S_z=\mbox{diag}\{\frac{N}{2},(\frac{N}{2}-1),...,-(\frac{N}{2}-1),-\frac{N}{2}\}$. Then we have $P^\dagger(\hat{\bk}\cdot {\bS})P=S_z$. Since $P$ rotates $\bS$ to the $z$-direction, it is easy to find $P=e^{-i\varphi S_z}e^{-i\theta S_y}$, which is a $(N+1)\times (N+1)$ matrix. Then we have the eigenvalues $E_i/(vk)=-N/2,-(N/2-1),...,N/2-1,N/2$, with the corresponding eigenstates $\Psi_i=P\cdot \bn_i$ where $\bn_i=(0,...,1,...,0)^T$, with all elements being 0 except for the $i$th which is 1.

\subsection{Pseudospin Hall effect}
The spectrum of $H$ has a $(N+1)$-fold degeneracy at $\bk=0$, which acts as a monopole in the momentum space. The Berry curvature emitted from the monopole exists in the whole space. From the semiclassical point of view, the motion of the wave packet is governed by the EOMs, while the Berry curvature induces an anomalous velocity when an effective force is exerted on the wave packet, resulting in a transverse shift\cite{Xiao2010}.
\begin{figure}[t]
  \centering
  \includegraphics[width=5cm]{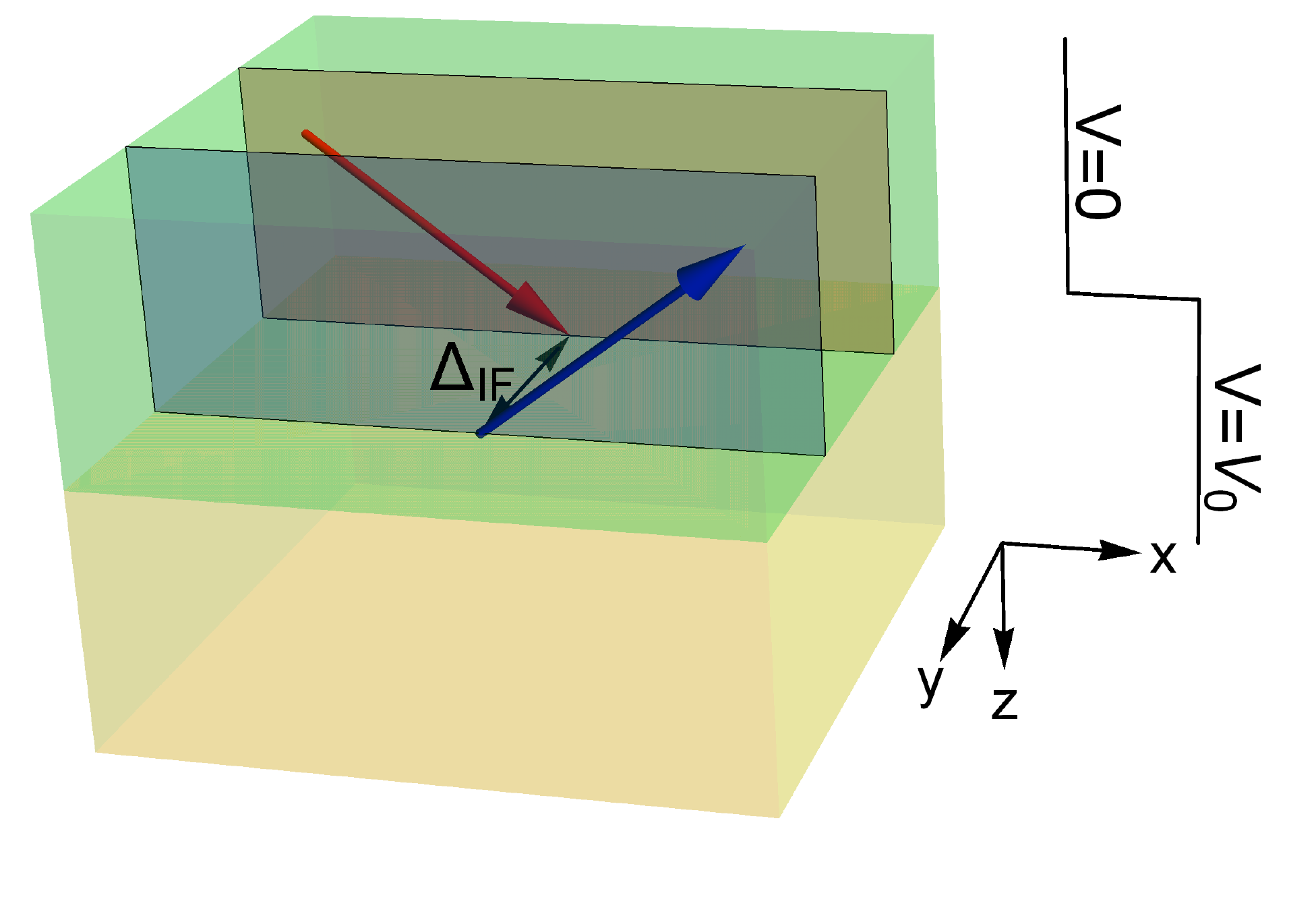}\label{IF_shift}
  \caption{The Imbert-Fedorov shift of topological fermions impinging on the interface between two topological semimetals.}\label{Fig1}
\end{figure}

The EOMs governing the trajectory of the real space coordinate $\br$ and the momentum space coordinate $\bk$ of the wave packet read\cite{Xiao2010}
\begin{eqnarray}
   \dot{\br}&=&\nabla_\bk E(\bk)-\dot{\bk}\times{\bf\Omega}(\bk), \label{eq:velocity}\\
   \dot{\bk}&=&-\nabla_\br U(\br)\label{eq:force},
\end{eqnarray}
where $\Omega(\bk)$ is the Berry curvature of a certain band. The potential $U(\br)$ can originate from an external electric field, or the interface with another material, or the surface confining potential. To be specific, we assume the wave packet is reflected at an interface, where a potential step is present, as shown in Fig.\ref{Fig1}. In principle, the potential step has to be smooth compared with the wave packet for the EOMs to hold; however, as previous studies have shown, whether the potential step is sharp or smooth does not affect the amount of the transverse shift\cite{Onoda2004,Jiang2015,Yang2015,Wang2017}. Moreover, the form of the potential step will not appear in our calculations. We assume the interface is at $z=0$, and the incident wave packet has momentum $(k_x^I,0,k_z^I)$. The IF shift $\Delta_{IF}$ is the integral of the anomalous velocity with respect to the time interval, during which the total reflection occurs,
\begin{eqnarray}
\Delta_{IF}=\int dt\,[{\dot{\bk}\times{\bf\Omega}^{(i)}(\bk)}]_y=\int_{k_z^I}^{-k_z^I} \Omega_x^{(i)} dk_z
\end{eqnarray}
where $\Omega_x^{(i)}=\frac{\partial{\Psi_i^{\dagger}}}{\partial{k_y}}\frac{\partial{\Psi_i}} {\partial{k_z}}-
\frac{\partial{\Psi_i^{\dagger}}}{\partial{k_z}}\frac{\partial{\Psi_i}}{\partial{k_y}}$.  For the topological fermions under consideration, $\Psi_i$ has been derived in the previous section. The integral can be written in the differential form\cite{He2012}
\begin{eqnarray}
\int \Omega_x^{(i)} dk_z&=&\int
dk_z\left(\frac{\partial{\Psi^{\dagger}_i}}{\partial{k_y}}\frac{\partial{\Psi_i}} {\partial{k_z}}-
\frac{\partial{\Psi^{\dagger}_i}}{\partial{k_z}}\frac{\partial{\Psi_i}} {\partial{k_y}}\right)\nonumber\\
&=&\int d\Psi^{\dagger}_i\wedge d\Psi_i,
\end{eqnarray}
where
\begin{eqnarray}
d\Psi_i&=&\left(-i S_ze^{-i\varphi S_z}e^{-i\theta S_y}d\varphi-ie^{-i\varphi S_z}e^{-i\theta S_y}S_yd\theta\right)\cdot \bn_i,\nonumber\\
d\Psi_i^{\dagger}&=&\bn_i^T\cdot\left(ie^{i\theta S_y}e^{i\varphi S_z}S_zd\varphi+iS_ye^{i\theta S_y}e^{i\varphi S_z}d\theta\right).
\end{eqnarray}
Then the IF shift becomes
\begin{eqnarray}
\Delta_{IF}
&=&\int \mbox{Tr}[e^{i\theta S_y}(S_yS_z-S_zS_y)e^{-i\theta S_y}\cdot(\bn_i\bn_i^T)]d\theta\wedge d\varphi\nonumber\\
&=&\int \mbox{Tr}[e^{i\theta S_y}S_xe^{-i\theta S_y}\cdot(\bn_i\bn_i^T)]d\theta\wedge d\varphi\nonumber\\
&=&\int \mbox{Tr}[(\cos\theta S_x+\sin\theta S_z)\cdot(\bn_i\bn_i^T)]d\theta\wedge d\varphi.
\end{eqnarray}
The diagonal elements of matrix $S_x$ are 0, thus
\begin{eqnarray}
\Delta_{IF}=(S_z)_{ii}\int \sin\theta d\theta\wedge d\varphi.
\end{eqnarray}
Since
\begin{eqnarray}
d\theta\wedge d\varphi=\frac{\partial{\theta}}{\partial{k_y}}\frac{\partial{\varphi}}{\partial{k_z}}-
\frac{\partial{\varphi}}{\partial{k_y}}\frac{\partial{\theta}}{\partial{k_z}},
\end{eqnarray}
where  $k_y=k\sin\theta\sin\varphi$ and $k_z=k\cos\theta$, it is straightforward to show
\begin{eqnarray}
&&\frac{\partial{\theta}}{\partial{k_y}}=\frac{1}{k\cos\theta\sin\varphi},\  \frac{\partial{\varphi}}{\partial{k_z}}=0, \nonumber\\ &&\frac{\partial{\varphi}}{\partial{k_y}}=\frac{1}{k\sin\theta\cos\varphi}, \ \frac{\partial{\theta}}{\partial{k_z}}=-\frac{1}{k\sin\theta},
\end{eqnarray}
hence
\begin{eqnarray}
d\theta\wedge d\varphi=\frac{1}{k^2\sin^2\theta\cos\varphi}
\end{eqnarray}
where $\cos\varphi=k_x/(k\sin\theta)$. Therefore,
\begin{eqnarray}
\Delta_{IF}=(S_z)_{ii}\int_{k_z^I}^{-k_z^I} \frac{1}{kk_x^I} dk_z=-(S_z)^2_{ii}\frac{2vk_z^I}{Ek_x^I}.
\end{eqnarray}
Finally, since the wave packet has $k_y=0$, we have $k_x^I/k_z^I=\tan\theta$, and the IF shift is expressed in terms of $\theta$,
\begin{eqnarray}
\Delta_{IF}=-(S_z)^2_{ii}\frac{2v\cot\theta}{E}.\label{eq:IF}
\end{eqnarray}

\begin{figure}[t]
  \centering
  \subfigure[]{\includegraphics[width=2.5cm]{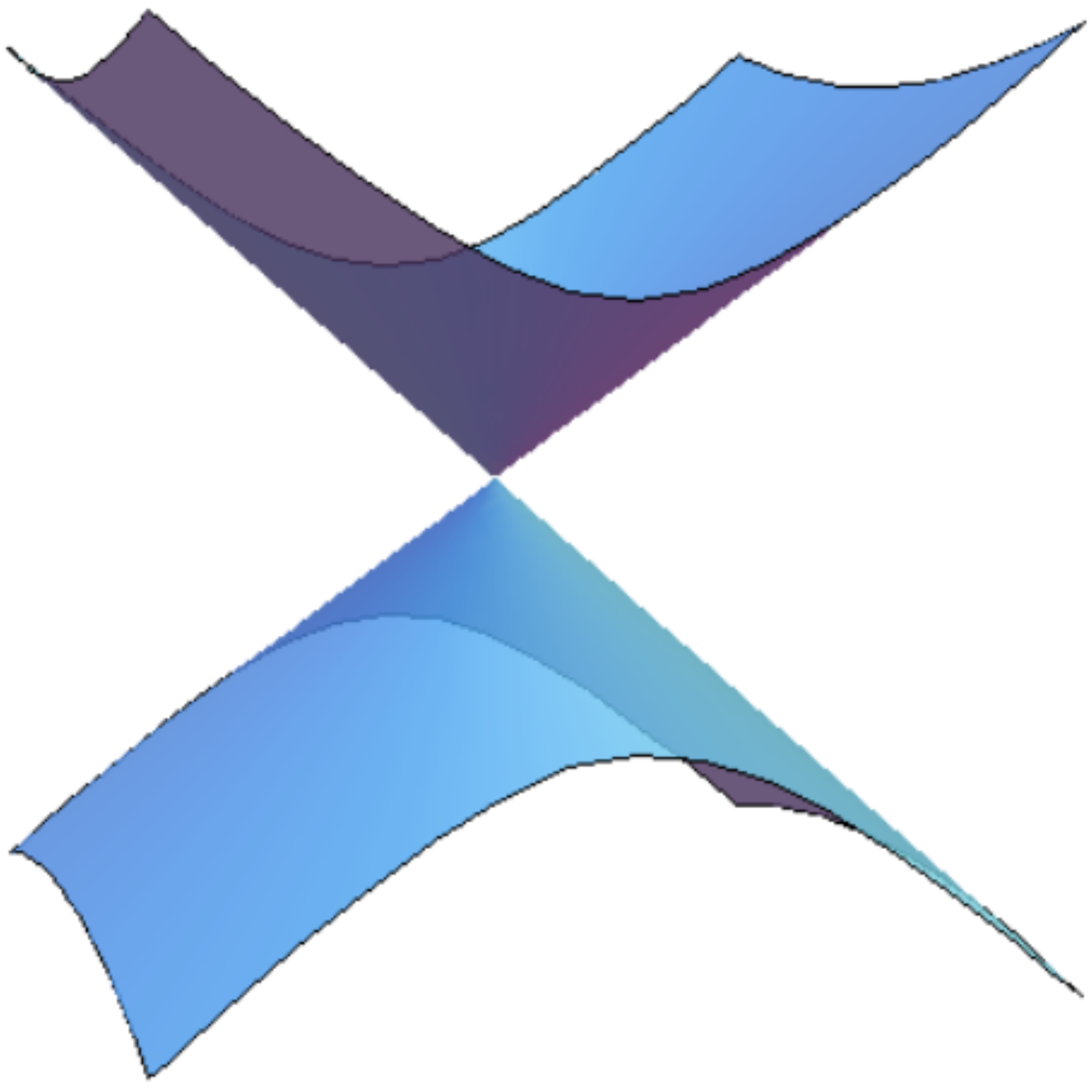}\label{spin1-2}}~~
  \subfigure[]{\includegraphics[width=4.5cm]{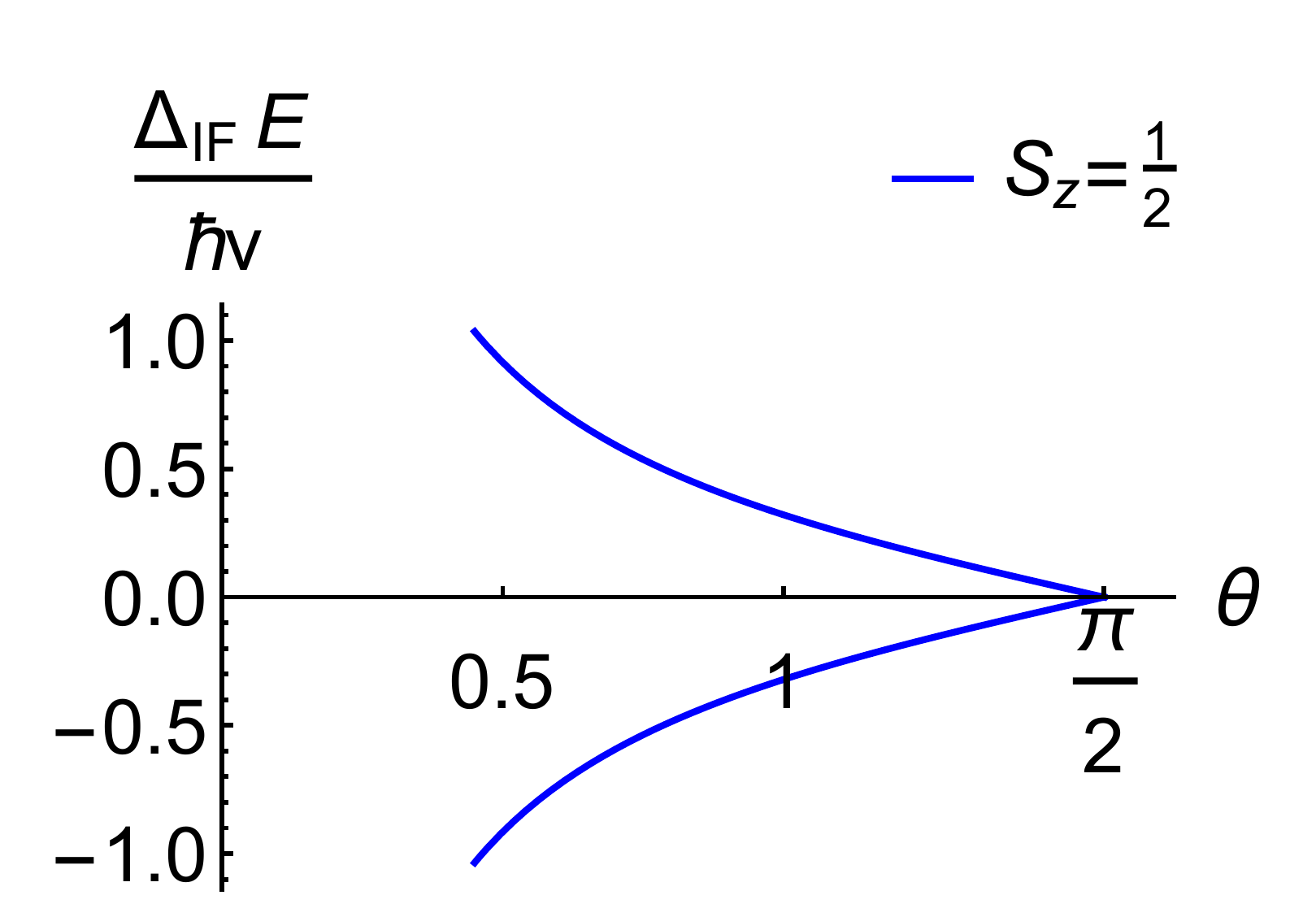}\label{IF1}}
  \subfigure[]{\includegraphics[width=2.5cm]{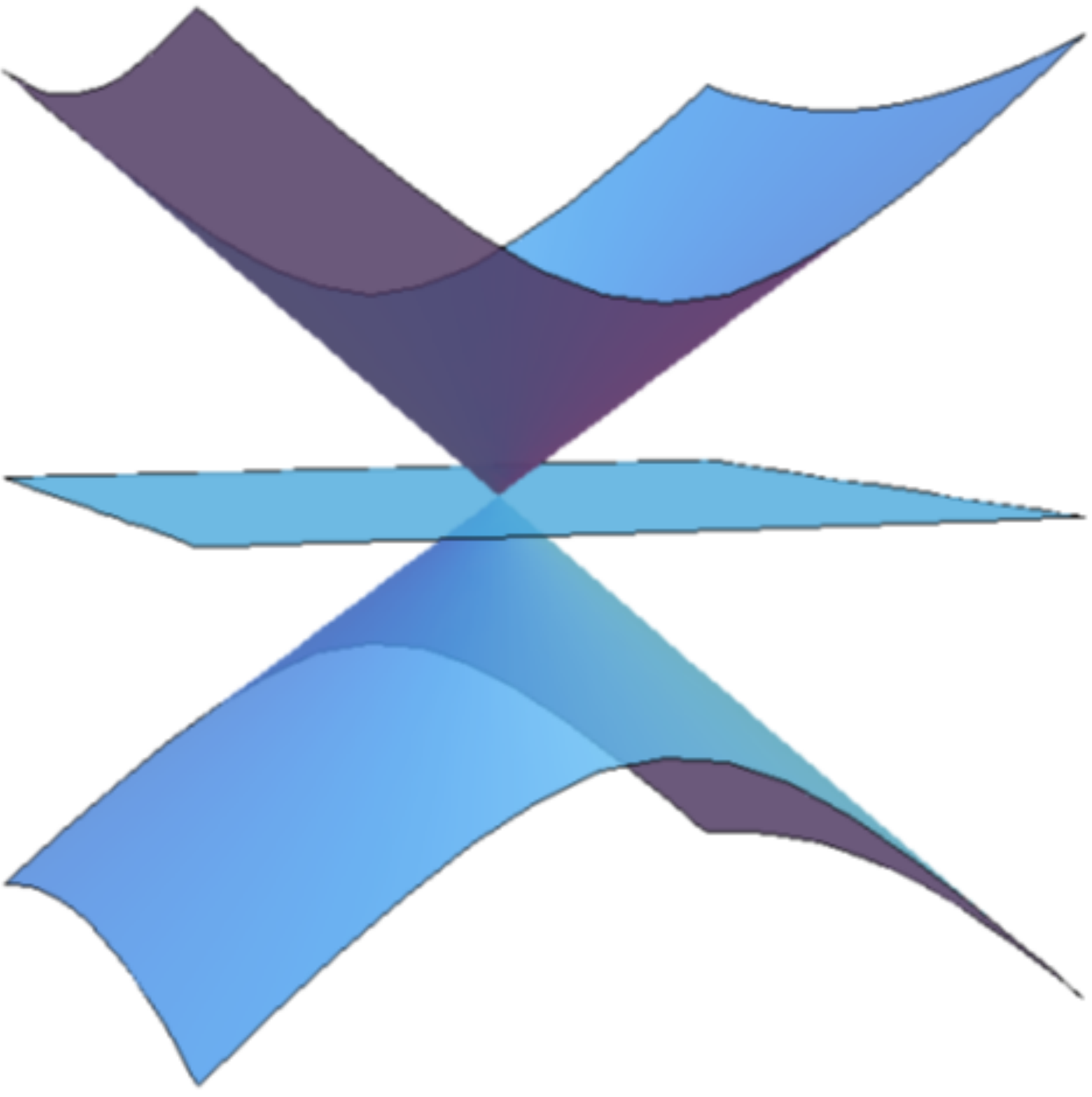}\label{spin1}}~~
  \subfigure[]{\includegraphics[width=4.5cm]{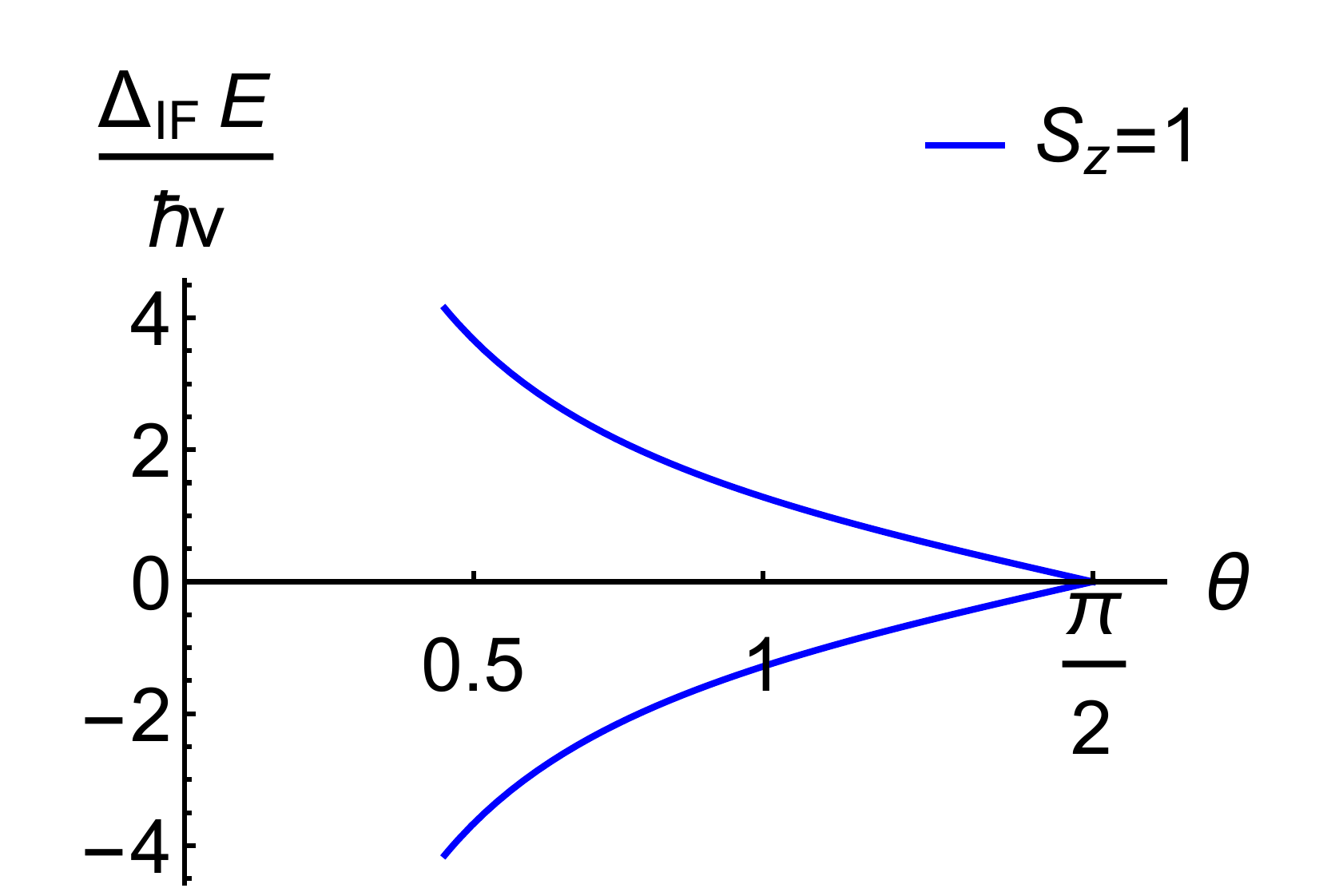}\label{IF2}}
  \subfigure[]{\includegraphics[width=2.5cm]{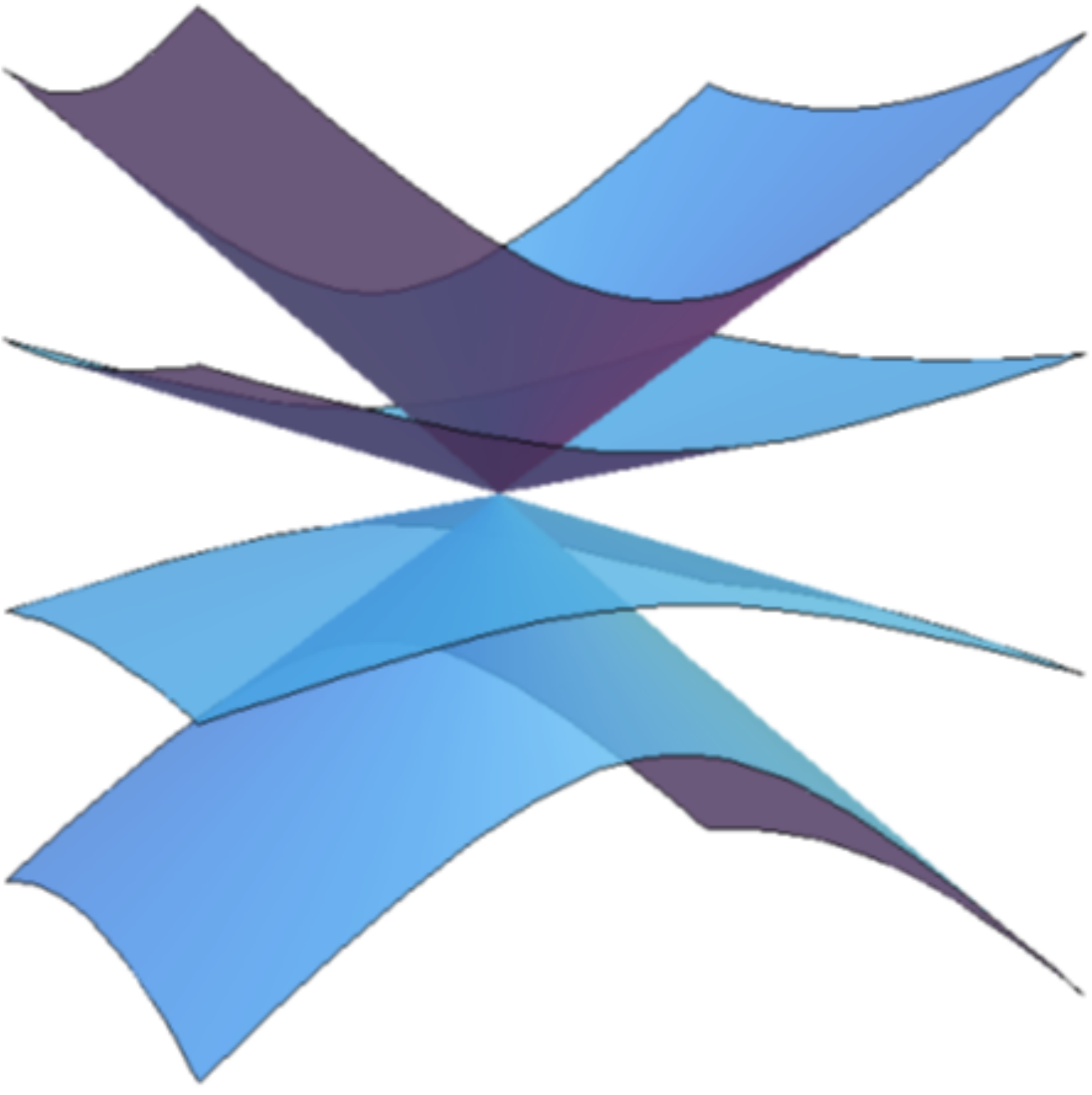}\label{spin3-2}}~~
  \subfigure[]{\includegraphics[width=4.5cm]{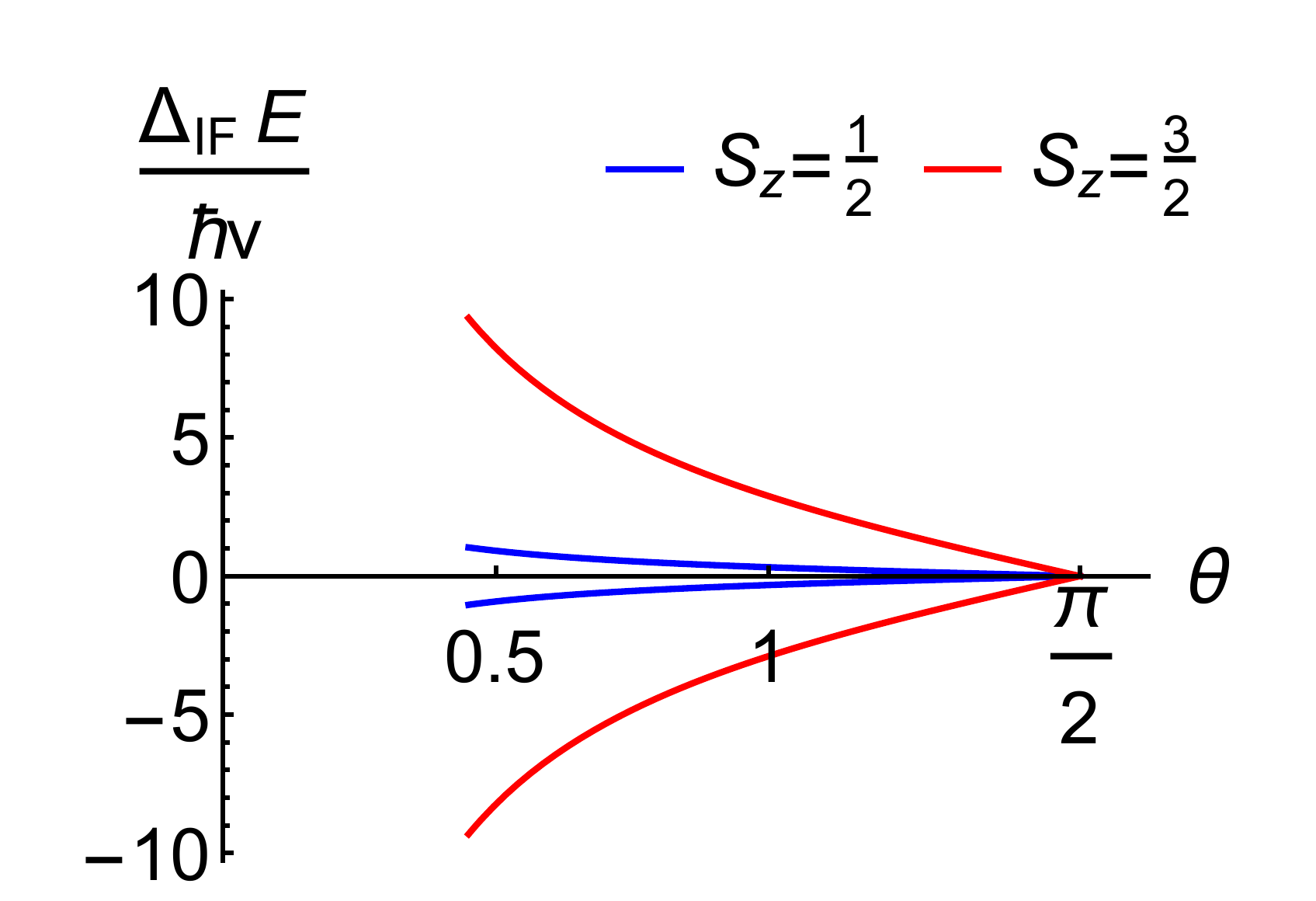}\label{IF3}}
  \caption{(a, c, e) Dispersion of pseudospin-1/2, pseudospin-1 and pseudospin-3/2 fermions, respectively. (b, d, f) The IF shift of pseudospin-1/2, pseudospin-1 and pseudospin-3/2 fermions, respectively.}\label{Fig2}
\end{figure}

The IF shift also depends on the chirality $C$\cite{Jiang2015,Yang2015}. If the chirality is reversed, e.g. $H=-v\bk\cdot\bS$, the direction of the IF shift is also reversed. Therefore,
\begin{eqnarray}
\Delta_{IF}=-C(S_z)^2_{ii}\frac{2v\cot\theta}{E}.
\end{eqnarray}
The results are shown in Fig.\ref{Fig2} for $N=1,2,3$, with $\hbar$ restored to make the variable dimensionless.
$N=1$ corresponds to WSMs, for which $\Delta_{IF}=-Cv\cot\theta/(2E)$. This seems to differ from the previous result by a factor of $\frac{1}{2}$\cite{Jiang2015,Yang2015}. This is due to the factor $\frac{1}{2}$ in $S_i=\frac{1}{2}\si_i$, and here $v$ is twice the group velocity of the Weyl fermions.  For $N=2$, we have $\Delta_{IF}=-2Cv\cot\theta/E$ for $S_z=\pm1$ components, while $\Delta_{IF}=0$ for $S_z=0$ component. This result can be applied straightforwardly to the Maxwell metal\cite{Zhu2017}. The pseudospin-3/2 semimetal needs special care since it is birefringent, or in other words, since it has two constant energy surfaces at the same energy with different group velocities. The fermions with $S_z=\pm\frac{3}{2}$ have IF shift $\Delta_{IF}=-9Cv\cot\theta/(2E)$, while for $S_z=\pm\frac{1}{2}$, $\Delta_{IF}=-Cv\cot\theta/(2E)$. In principle, the fermions with $S_z=\frac{3}{2}$ could tunnel into the band with $S_z=\frac{1}{2}$ and vice versa\cite{Wang2017}, resulting in a different IF shift. A lattice model with a band cutoff is required to study the quantum tunneling, which is beyond the scope of this work.

\begin{figure}[t]
  \centering
  \subfigure[]{\includegraphics[width=3.cm]{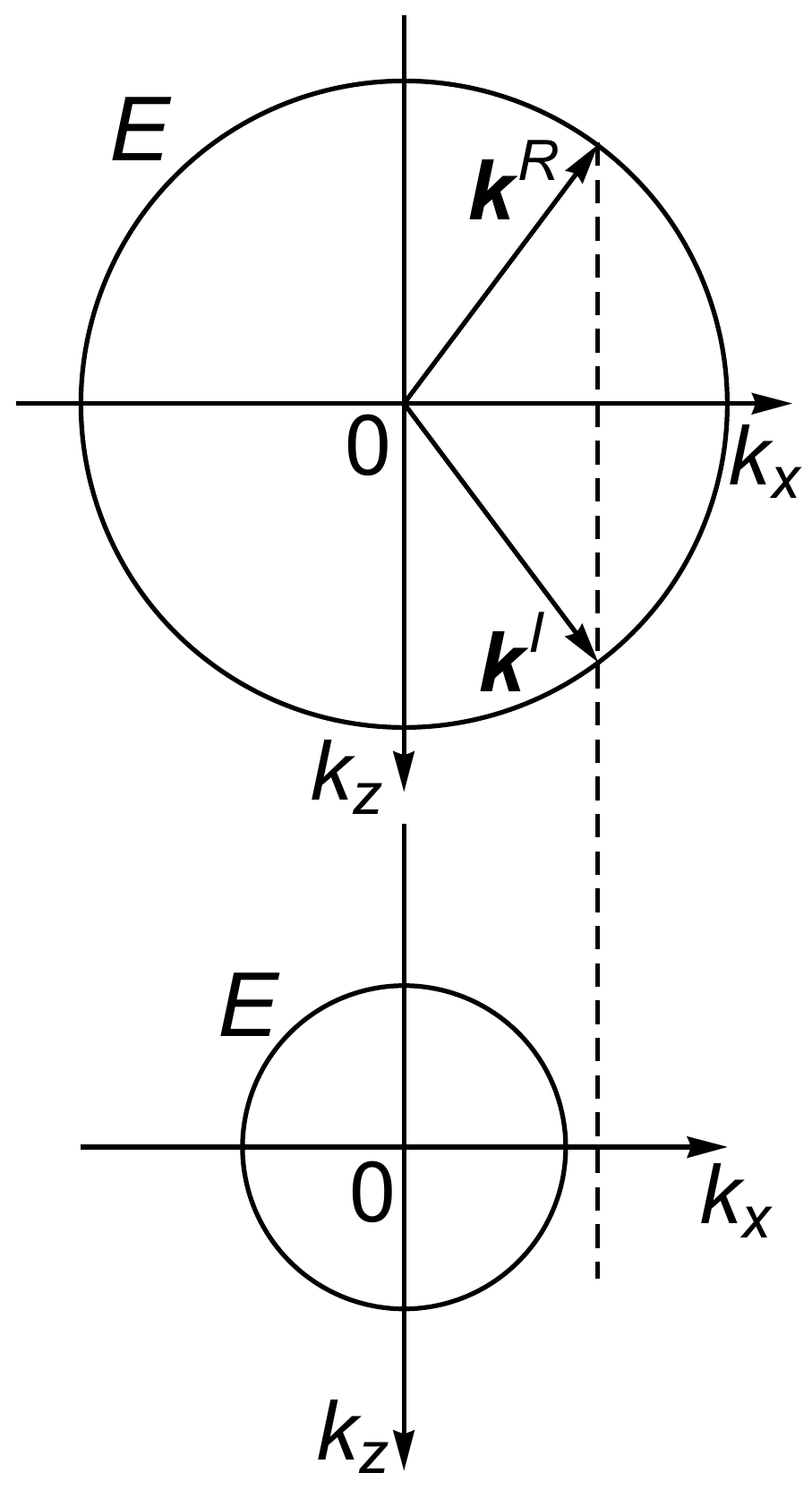}\label{tot_ref}}~~~~
  \subfigure[]{\includegraphics[width=4.cm]{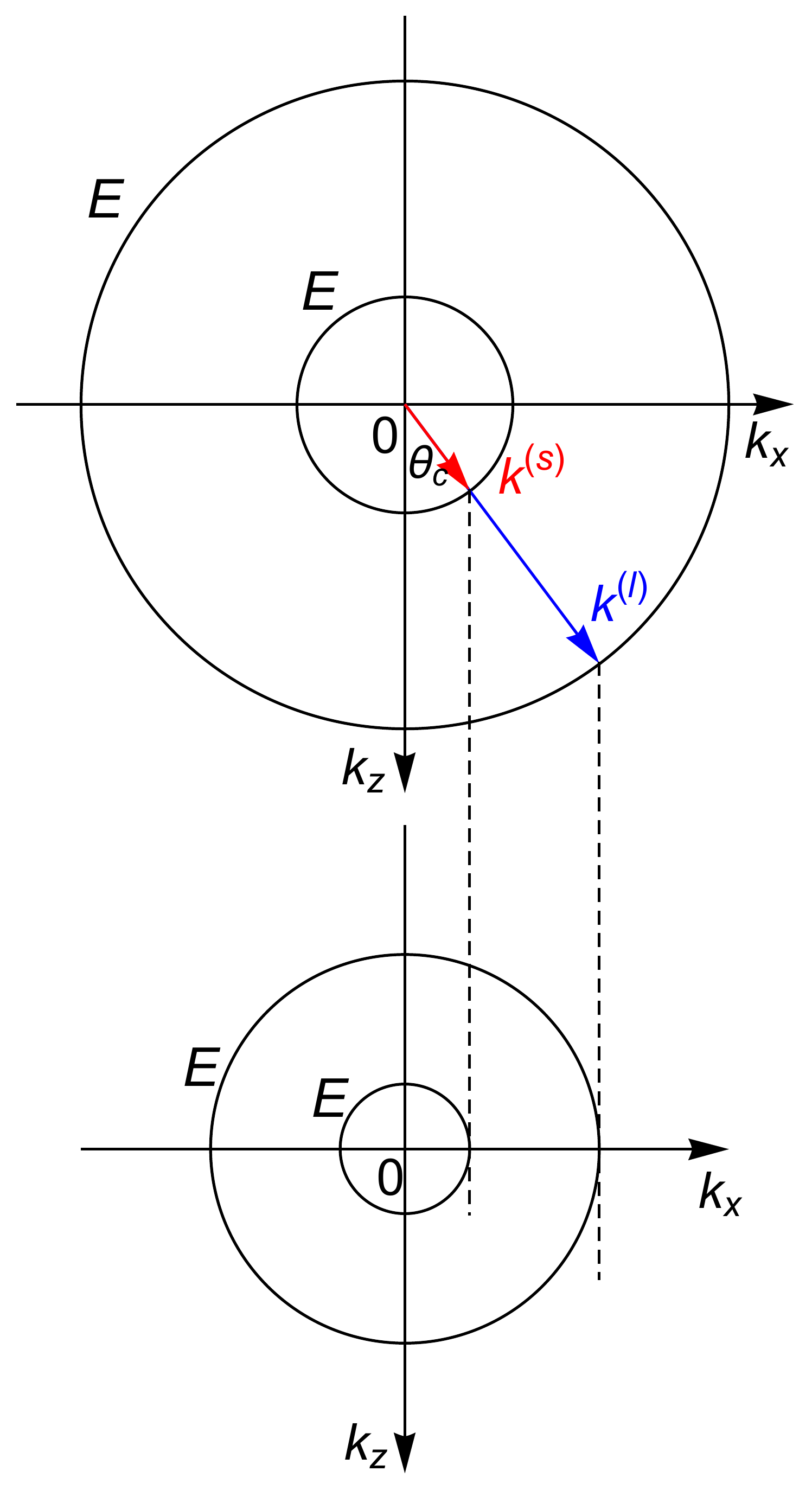}\label{crit_ang}}
  \caption{(a) Total reflection occurs when the angle of incidence is above the critical angle. (b) The critical angles of pseudospin-3/2 fermions with different $S_z$ components are the same.}\label{Fig3}
\end{figure}

Now we clarify the condition for total reflection to occur. During the process shown in Fig.\ref{Fig1}, the energy $E$ and $k_x$ are conserved, and $k_y=0$. We plot the contours with energy $E$ in $k_x-k_z$ plane for the two topological semimetals in Fig.\ref{tot_ref}. As long as there is no available state in the lower semimetal to receive the incident fermions from the upper one, i.e. the incident momentum $\bk^I$'s projection on the $k_x$-axis is larger than the radius of the lower contour, the fermions are totally reflected. Then the critical angle is readily derived, $\theta_c=\sin^{-1}(k^A/k^B)$, where $k^A(k^B)$ denotes the radius of the constant energy contour of the upper(lower) semimetal in Fig.\ref{Fig1}. For the pseudospin-3/2 semimetal, the fermions with $S_z=\frac{3}{2}$ and $\frac{1}{2}$ have the same critical angle as shown in Fig.\ref{crit_ang}, because the ratio of the larger and smaller radii $k^{(l)}/k^{(s)}=3$ for any energy $E$. This only holds for linear dispersion.

\subsection{Conservation of the total angular momentum}
In WSMs, the IF shift can be understood in the way of the conservation of the total angular momentum (TAM)\cite{Yang2015,Wang2017}. Here, a similar interpretation can be rendered. The TAM is defined as $\bJ=\bL+\bS$, where $\bL$ is the orbital angular momentum operator. In the setup of Fig.\ref{Fig1}, the $z$-component of the TAM, $J_z=L_z+S_z$, is conserved, i.e. $[J_z,H_C]=0$, where $H_C$ is the Hamiltonian of the composite system with both semimetals. Therefore, $\langle J_z^I\rangle=\langle J_z^R\rangle$, with $I(R)$ labels the incident(reflected) fermions. Since $\langle L_z\rangle=xk_y-yk_x$ and $k_y=0$, we have the transverse shift
\begin{eqnarray}
\Delta_{IF}=y^{R}-y^{I}=\frac{1}{k_x^I}\left(\langle S_{z}^{R}\rangle-\langle S_{z}^{I}\rangle\right).
\end{eqnarray}
The expectation value of $S_z$ in the $i$th band is
\begin{eqnarray}
  \langle S_z\rangle_i&=&\langle\Psi_i|S_z|\Psi_i\rangle\nonumber\\
  &=&\langle\bn_i|P^\dagger S_zP|\bn_i\rangle
\end{eqnarray}
Since $e^{i\theta S_y}S_ze^{-i\theta S_y}=S_z\cos\theta-S_x\sin\theta$, then
\begin{eqnarray}
  \langle S_z\rangle_i&=&\langle\bn_i|S_z\cos\theta-S_x\sin\theta|\bn_i\rangle =(S_z)_{ii}\cos\theta.
\end{eqnarray}
If the angle of incidence is $\theta$, then the angle of the reflected beam is $\pi-\theta$, therefore, $\langle S_{z}^{R}\rangle-\langle S_{z}^{I}\rangle=-2(S_z)_{ii}\cos\theta$, and we recover the result Eq.\ref{eq:IF}.

\section{The IF shift in nodal-line semimetals}\label{SecIII}
NLSMs host one-dimensional curves in the Brillouin zone along which the conduction and the valence bands cross each other. The classification of NLSMs is not complete yet. So far, it has been found that nodal lines can be protected by one or several of these spatial symmetries: inversion, mirror reflection and twofold screw rotation symmetries, and they belong to different topological classes according to their symmetries\cite{Fang2016}. Since different classes have different low energy effective Hamiltonians, the form of the Berry curvature and hence the IF shift should also be different. We study the IF shift in two types of NLSMs, one protected by both time-reversal and inversion symmetries and the other, the so-called vortex ring model, by mirror reflection symmetry.

\subsection{Vanishing IF shift in NLSM with time-reversal and inversion symmetries}
The model of NLSM protected by time-reversal and inversion symmetries is given by $H_{NL}=(k_x^2+k_y^2-k_0^2)\si_x+k_z\si_y$\cite{Fang2016}. It is straightforward to show the Berry curvature ${\bf\Omega}(\bk)$ is zero. Actually, the time-reversal symmetry requires ${\bf\Omega}(\bk)=-{\bf\Omega}(-\bk)$, and the inversion symmetry requires ${\bf\Omega}(\bk)={\bf\Omega}(-\bk)$, thus ${\bf\Omega}(\bk)=0$ and the IF shift vanishes.

If the reflection occurs at the $z=0$ plane, the vanishing of the IF shift can be seen from the conservation of the {\it orbital} angular momentum. Since $[L_z,H_{NL}]=0$, $\langle L_z\rangle=xk_y-yk_x$ is a conserved quantity. Assume $k_y=0$, then since $k_x$ is conserved, $y$ is also unchanged during the reflection.

\begin{figure}[t]
  \centering
  \subfigure[]{\includegraphics[width=4.cm]{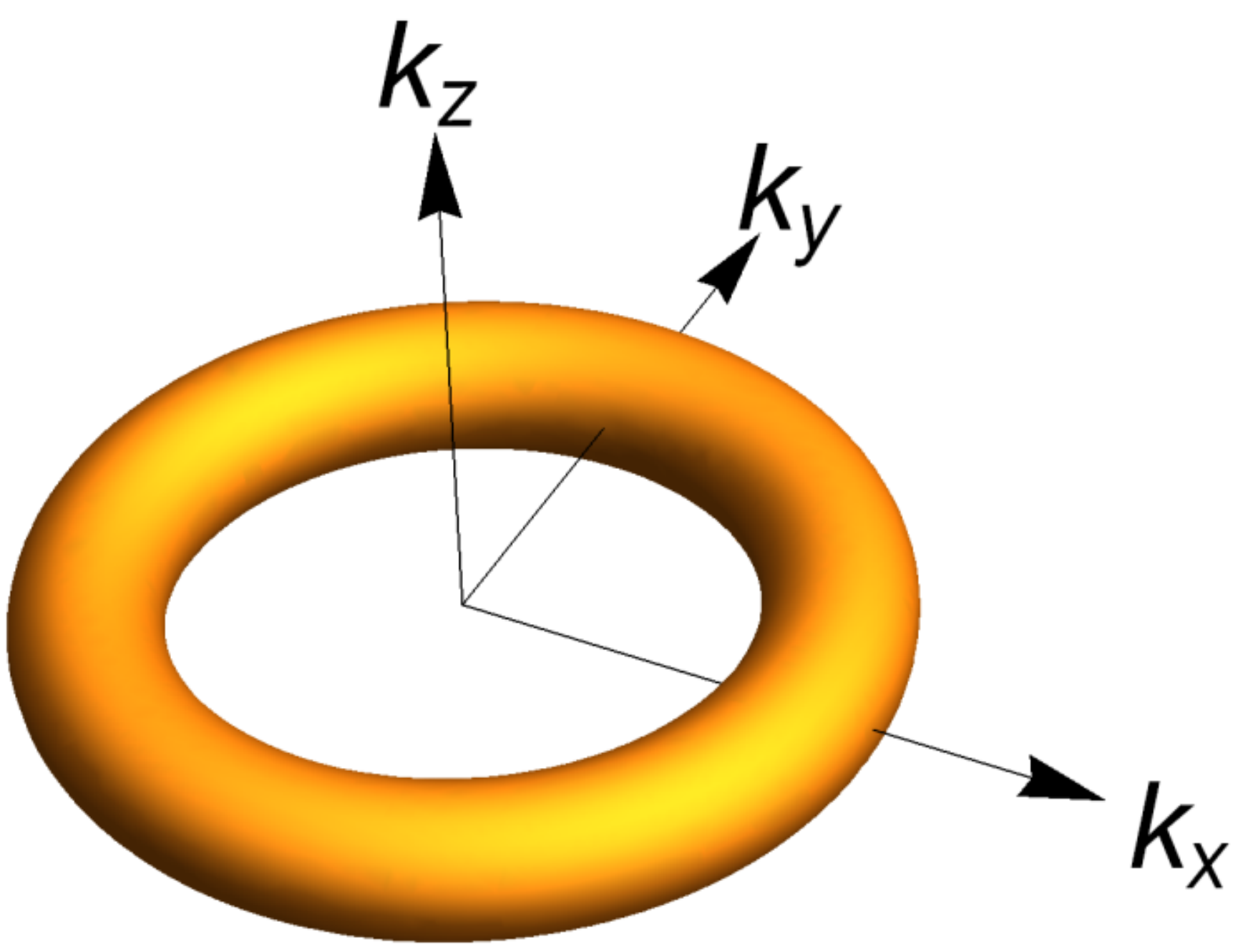}\label{NL1}}~~
  \subfigure[]{\includegraphics[width=4.cm]{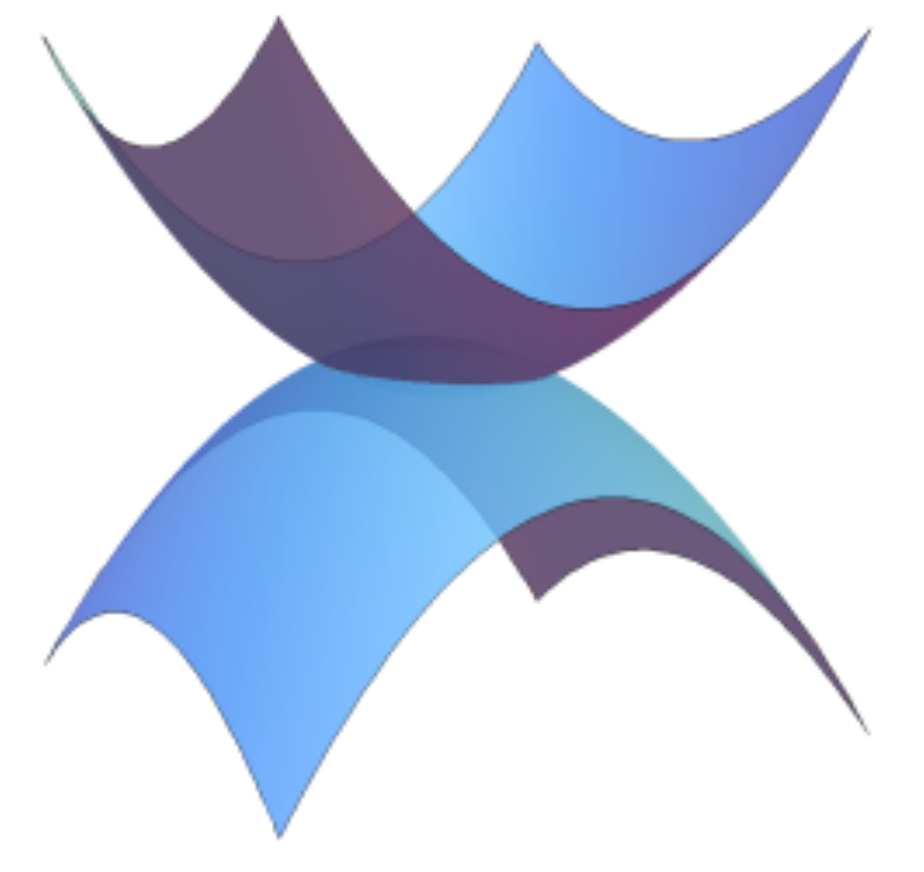}\label{NL2}}
  \subfigure[]{\includegraphics[width=4.cm]{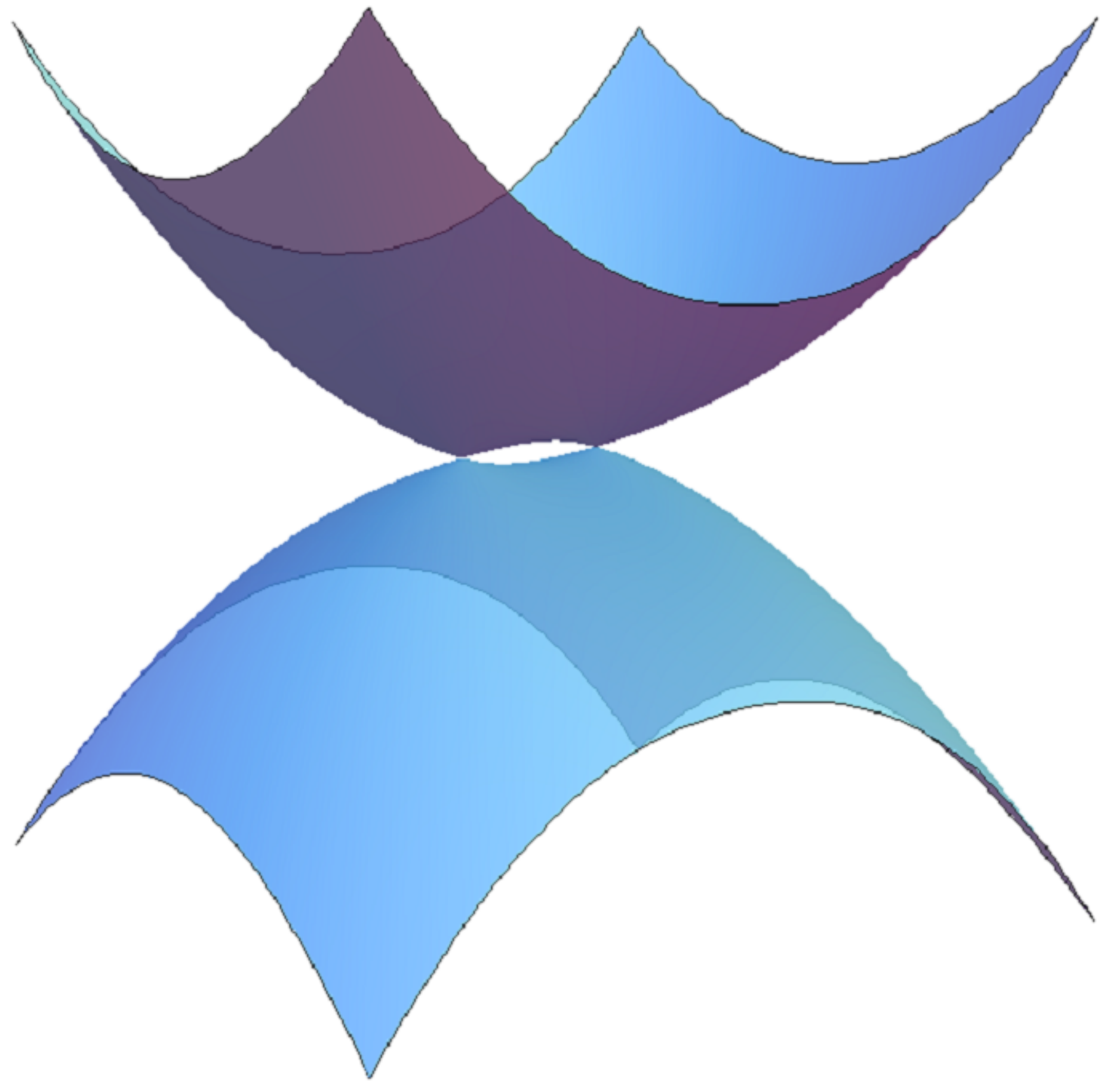}\label{NL3}}~~
  \subfigure[]{\includegraphics[width=4.cm]{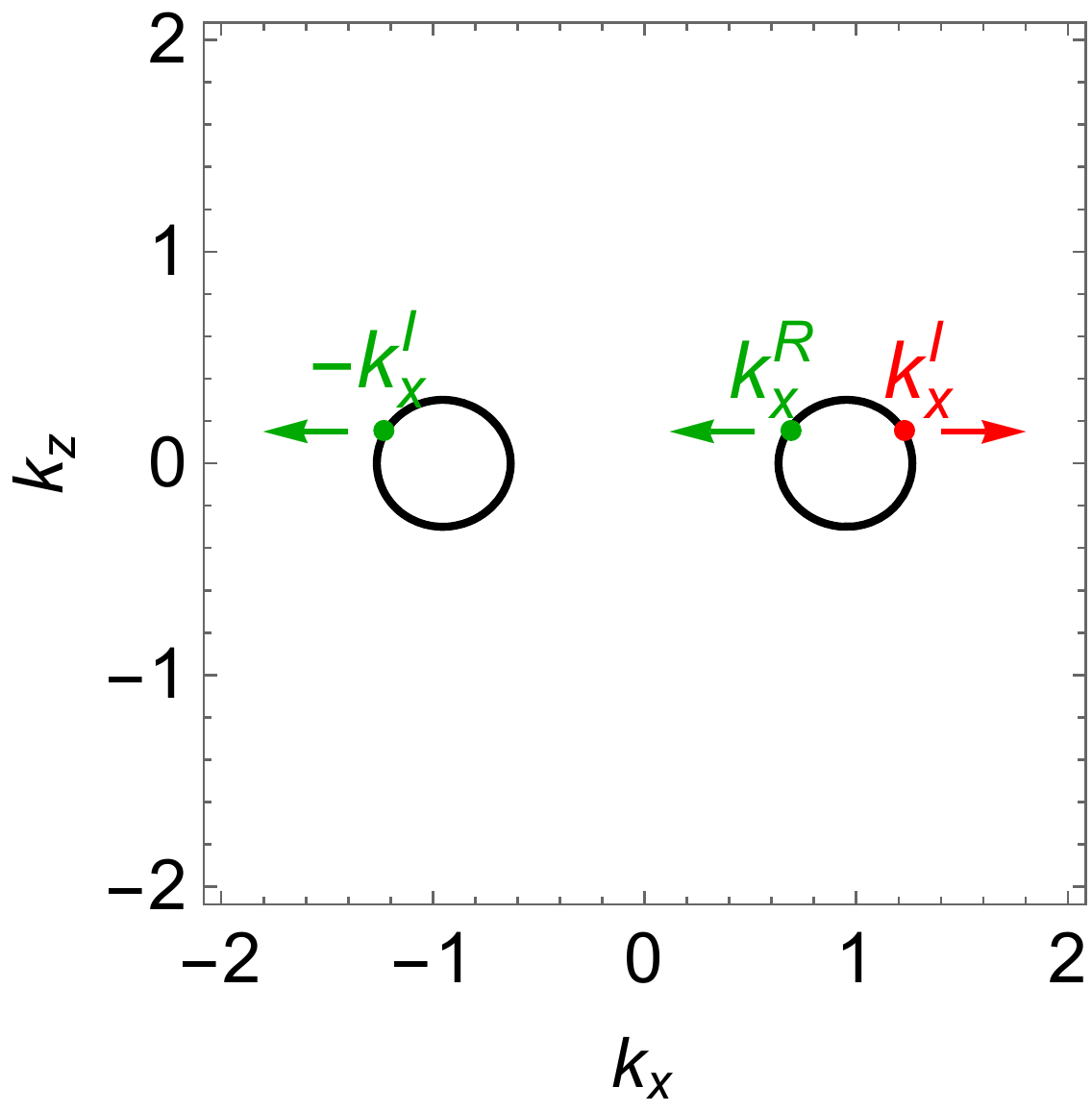}\label{NL4}}
  \caption{(a) The toroidal Fermi surface close to the nodal ring. (b) The spectrum for $k_z=0$. (c) The spectrum for $k_y=0.8$. (d) Schematic of the two possibilities of reflection process.} \label{Fig_NL}
\end{figure}
\subsection{The IF shift in NLSM with a vortex ring}
The vortex ring model reads
\begin{eqnarray}\label{eq:Hvr}
  H_{VR}(\bk) &=& -\frac{1}{m_z}k_xk_z\si_x-\frac{1}{m_z}k_yk_z\si_y\nonumber\\
  &&+\frac{1}{2m_r}\left(k_x^2+k_y^2-k_z^2-k_0^2\right)\si_z,
\end{eqnarray}
which breaks both time-reversal and inversion symmetry, but has mirror reflection symmetry\cite{Lim2017}. The energy spectrum is given by
\begin{eqnarray}
  E_\pm &=& \pm\sqrt{(k_r^2-k_z^2-k_0^2)^2/(4m_r^2)+k_r^2k_z^2/m_z^2}
\end{eqnarray}
where $k_r\equiv\sqrt{k_x^2+k_y^2}$. The nodal ring is given by $k_z=0$, $k_r=k_0$. At low energy, $E_\pm$ has the form $E_\pm\approx\pm k_0 \sqrt{(k_r-k_0)^2/m_r^2+k_z^2/m_z^2}$. The low energy Fermi surface is a torus with an elliptical cross section, as shown in Fig.\ref{NL1}.
At a fixed energy $E$, we can parameterize $\bk$ by $(E,\theta,\phi)$ as
\begin{eqnarray}
  k_x &=& (\frac{E m_r\sin\theta}{k_0}+k_0)\cos\phi,\\
  k_y &=& (\frac{E m_r\sin\theta}{k_0}+k_0)\sin\phi,\\
  k_z &=& \frac{E m_z\cos\theta}{k_0},
\end{eqnarray}
and the velocity $\bv=\nabla_\bk E_\pm$ at low energy has components
\begin{eqnarray}
  v_x &=& \pm\frac{k_0}{m_r}\sin\theta\cos\phi,\\
  v_y &=& \pm\frac{k_0}{m_r}\sin\theta\sin\phi,\\
  v_z &=& \pm\frac{k_0}{m_z}\cos\theta.
\end{eqnarray}
Unlike the isotropic system studied in the previous section, the NLSM with a vortex ring is highly anisotropic. For total reflection, we anticipate that the IF shifts are distinct for the reflection surfaces in different directions. Considering that the $x$- and $y$-directions are equivalent in this model, we study two cases, with the reflection surface (i) at the $z=0$ plane and (ii) at the $x=0$ plane, assuming the incident electrons are at low energy in the upper band. In case (i), during reflection, the momentum in $z$ direction, $k_z$, is reversed, while $k_x$ and $k_y$ are conserved. Therefore, the IF shift is the vector $(-\int_{k_z^I}^{-k_z^I} dk_z \Omega_y, \int_{k_z^I}^{-k_z^I} dk_z \Omega_x)$ in the $xy$ plane projected to the direction perpendicular to the $(v_x,v_y)$. For the upper band, the three components of the Berry curvature are given by
\begin{eqnarray}\label{eq:BCx}
  \Omega_x &=& -\frac{k_xk_z(k_r^2+k_z^2-k_0^2)}{4m_rm_z^2E_+^3},\\
  \Omega_y &=& -\frac{k_yk_z(k_r^2+k_z^2-k_0^2)}{4m_rm_z^2E_+^3},\\
  \Omega_z &=& -\frac{k_z^2(k_r^2+k_z^2+k_0^2)}{4m_rm_z^2E_+^3}.\label{eq:BCz}
\end{eqnarray}
Noticing that $\Omega_x$ and $\Omega_y$ are odd functions of $k_z$, we conclude that $\Delta_{IF}=0$ in case (i). In case (ii), $k_x$ is not conserved, while $k_y$, $k_z$ and $E$ are the same in the initial and final state. One possibility is $\phi\rightarrow \pi-\phi$ while fixing $\theta$, corresponding to $k_x^I\rightarrow-k_x^I$; another possibility is $\theta\rightarrow-\theta$ and $\phi\rightarrow\sin^{-1}[(\frac{E m_r}{k_0}\sin\theta+k_0)\sin\phi/(-\frac{E m_r}{k_0}\sin\theta+k_0)]$, corresponding to $k_x^I\rightarrow k_x'$, as shown in Fig.\ref{NL4}, and we use $\Delta_{IF}^{(1)}$ and $\Delta_{IF}^{(2)}$ to denote them, respectively. It is straightforward to show that $\Delta_{IF}(E,\theta,\phi)= -\Delta_{IF}(E,\pi-\theta,\phi)=-\Delta_{IF}(E,\theta,\pi-\phi)=\Delta_{IF}(E,\theta,-\phi)$, so we need only consider $\Delta_{IF}$ in the range $\{0,\frac{\pi}{2}\}$. Formally,
\begin{eqnarray}
  \Delta_{IF}^{(1)} &=& (-\int_{k_x^I}^{-k_x^I}dk_x\Omega_z,\int_{k_x^I}^{-k_x^I}dk_x\Omega_y)\cdot\frac{(-v_z,v_y)} {\sqrt{v_y^2+v_z^2}},\\
  \Delta_{IF}^{(2)} &=& (-\int_{k_x^I}^{k_x'}dk_x\Omega_z,\int_{k_x^I}^{k_x'}dk_x\Omega_y)\cdot\frac{(-v_z,v_y)} {\sqrt{v_y^2+v_z^2}},
\end{eqnarray}

\begin{figure}[t]
  \centering
  \subfigure[]{\includegraphics[width=4.cm]{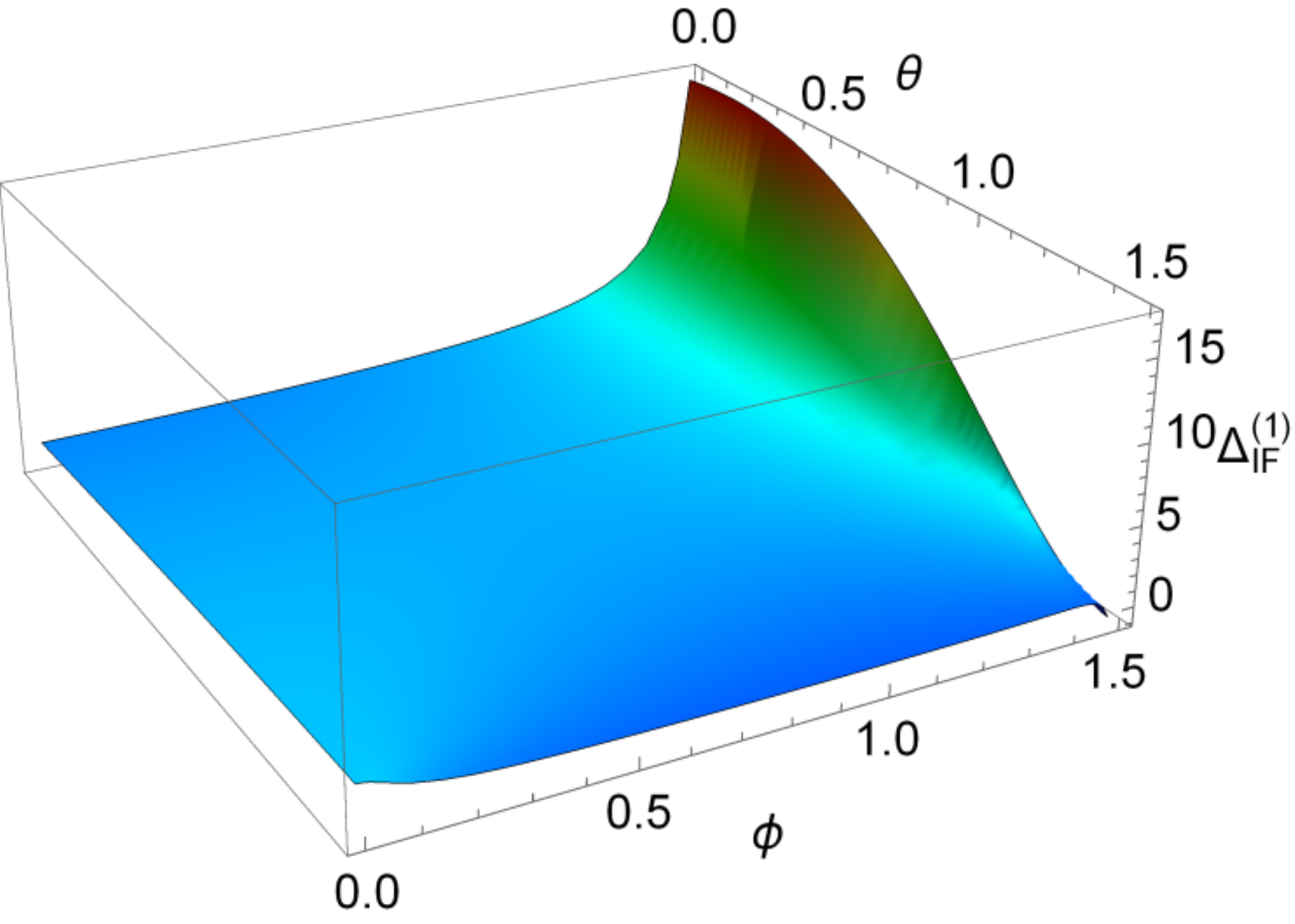}\label{IF_NL1}}~~
  \subfigure[]{\includegraphics[width=4.cm]{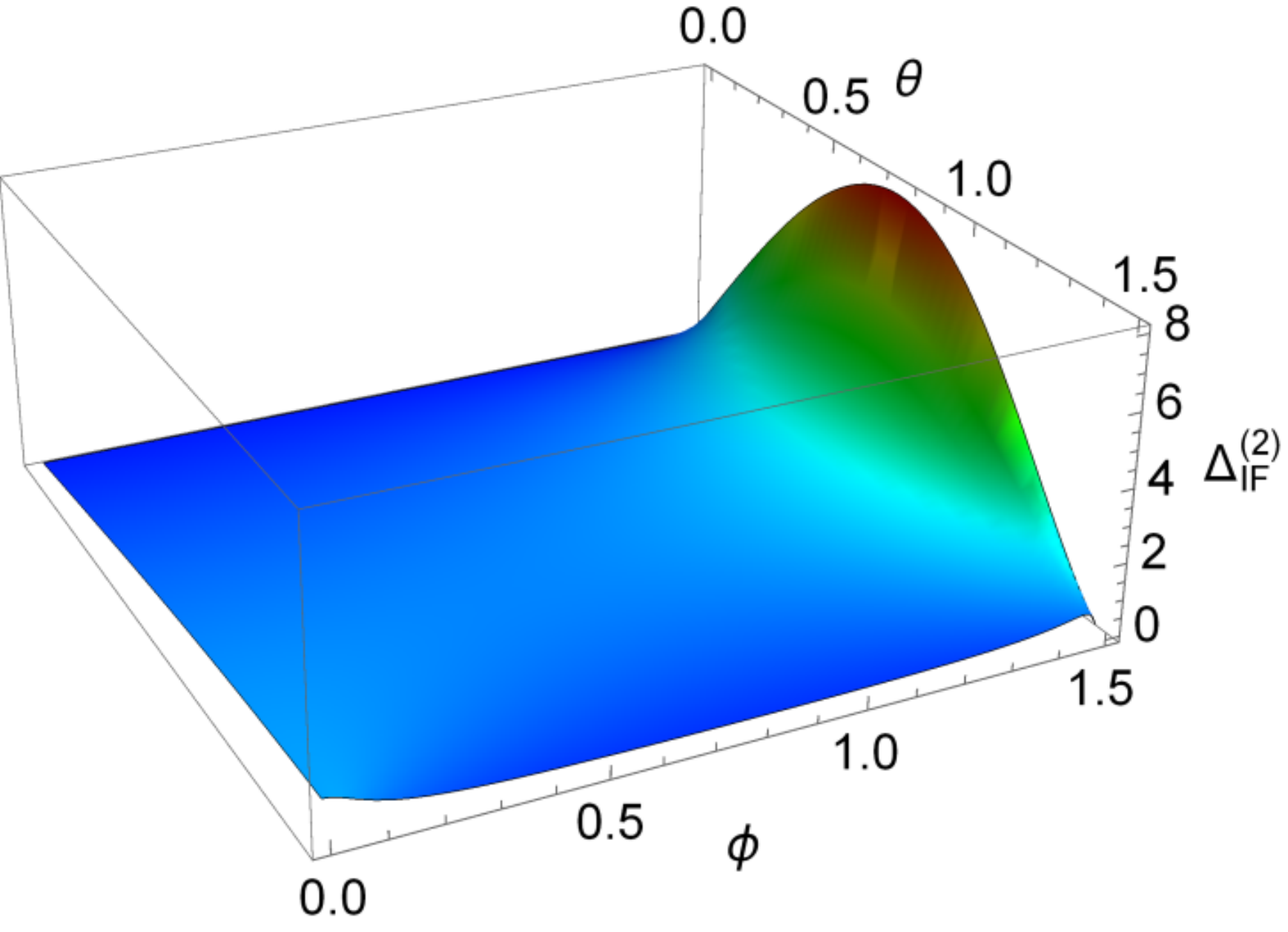}\label{IF_NL2}}
  \subfigure[]{\includegraphics[width=4.cm]{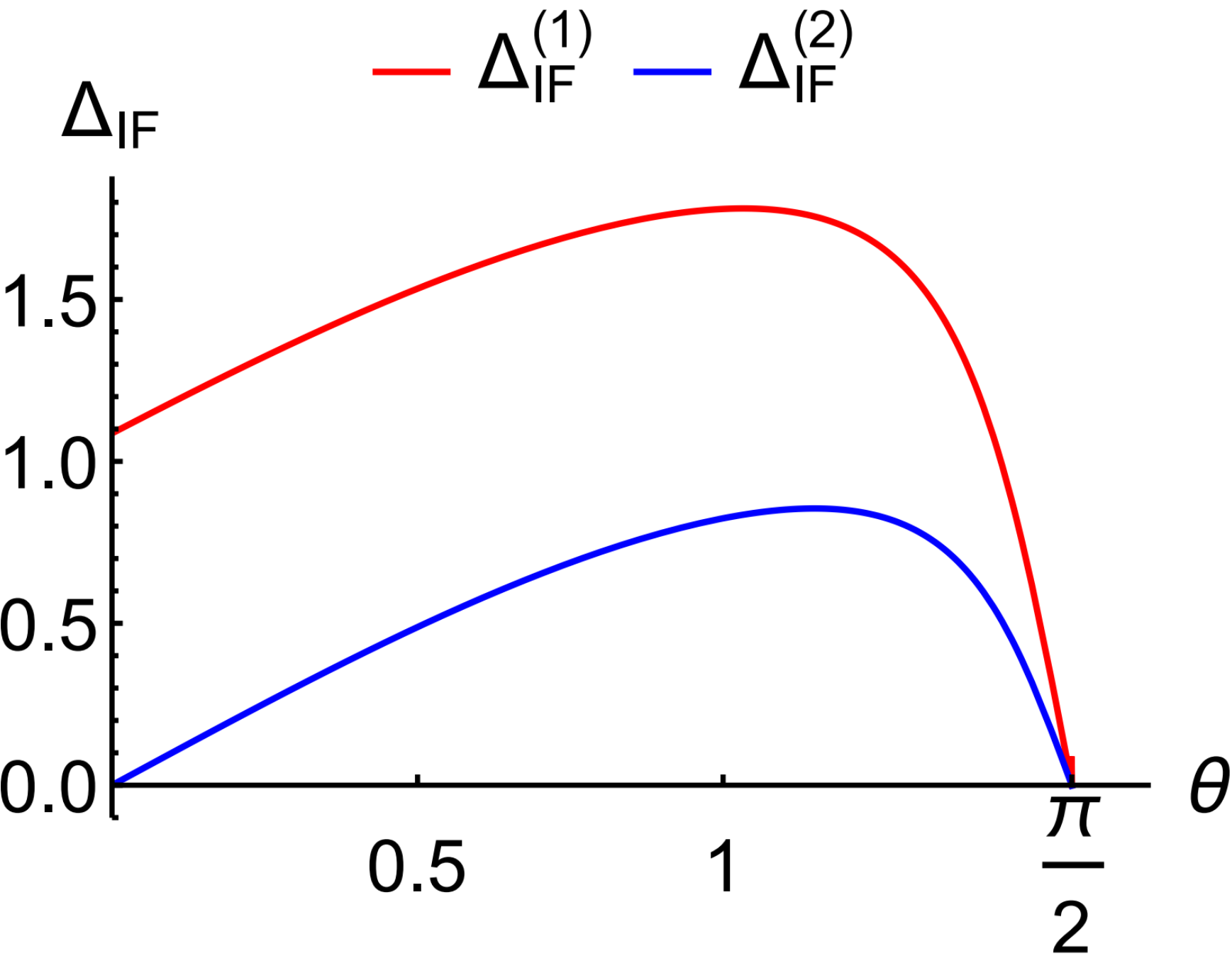}\label{IF_NL3}}~~
  \subfigure[]{\includegraphics[width=4.cm]{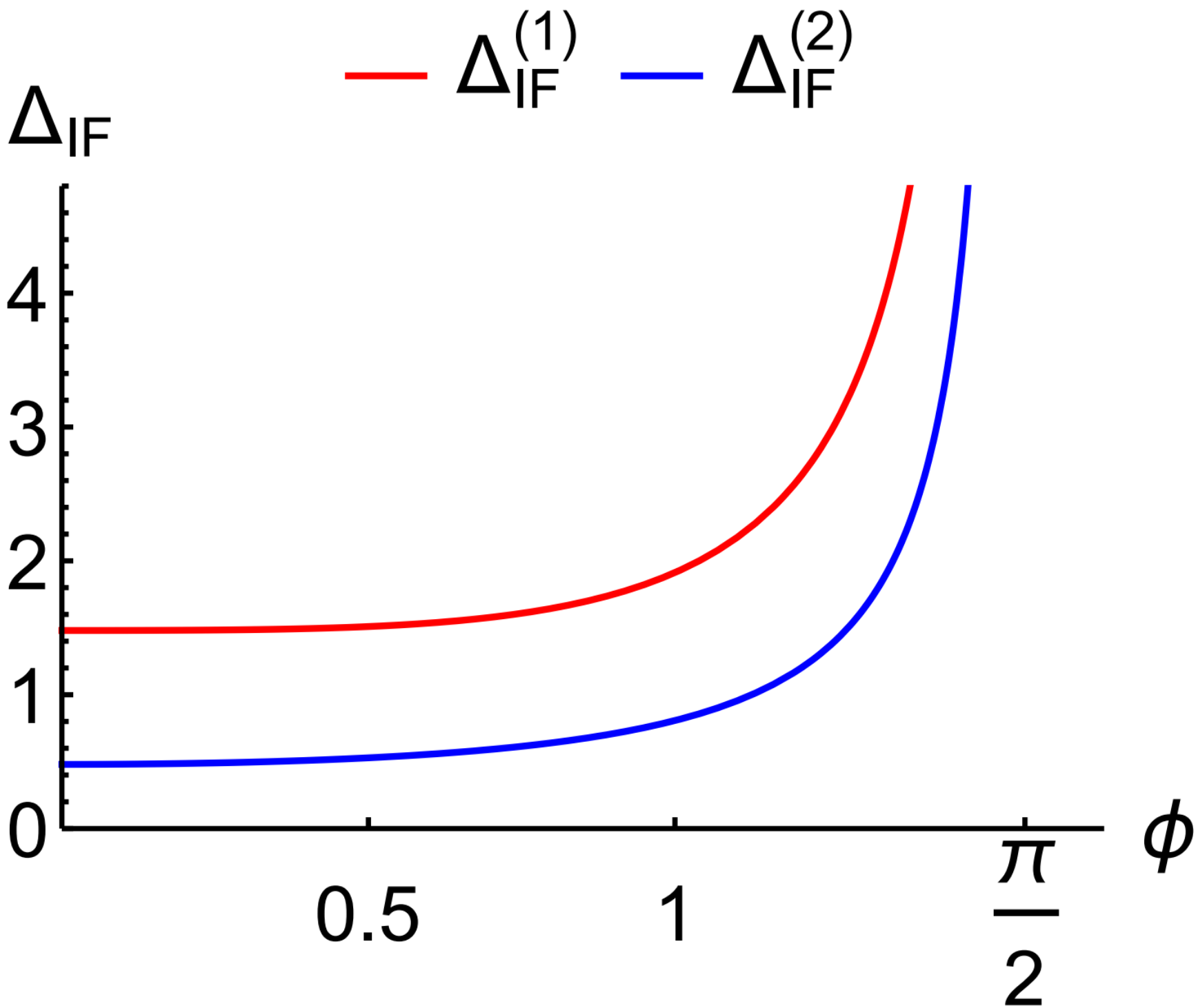}\label{IF_NL4}}
  \subfigure[]{\includegraphics[width=4.cm]{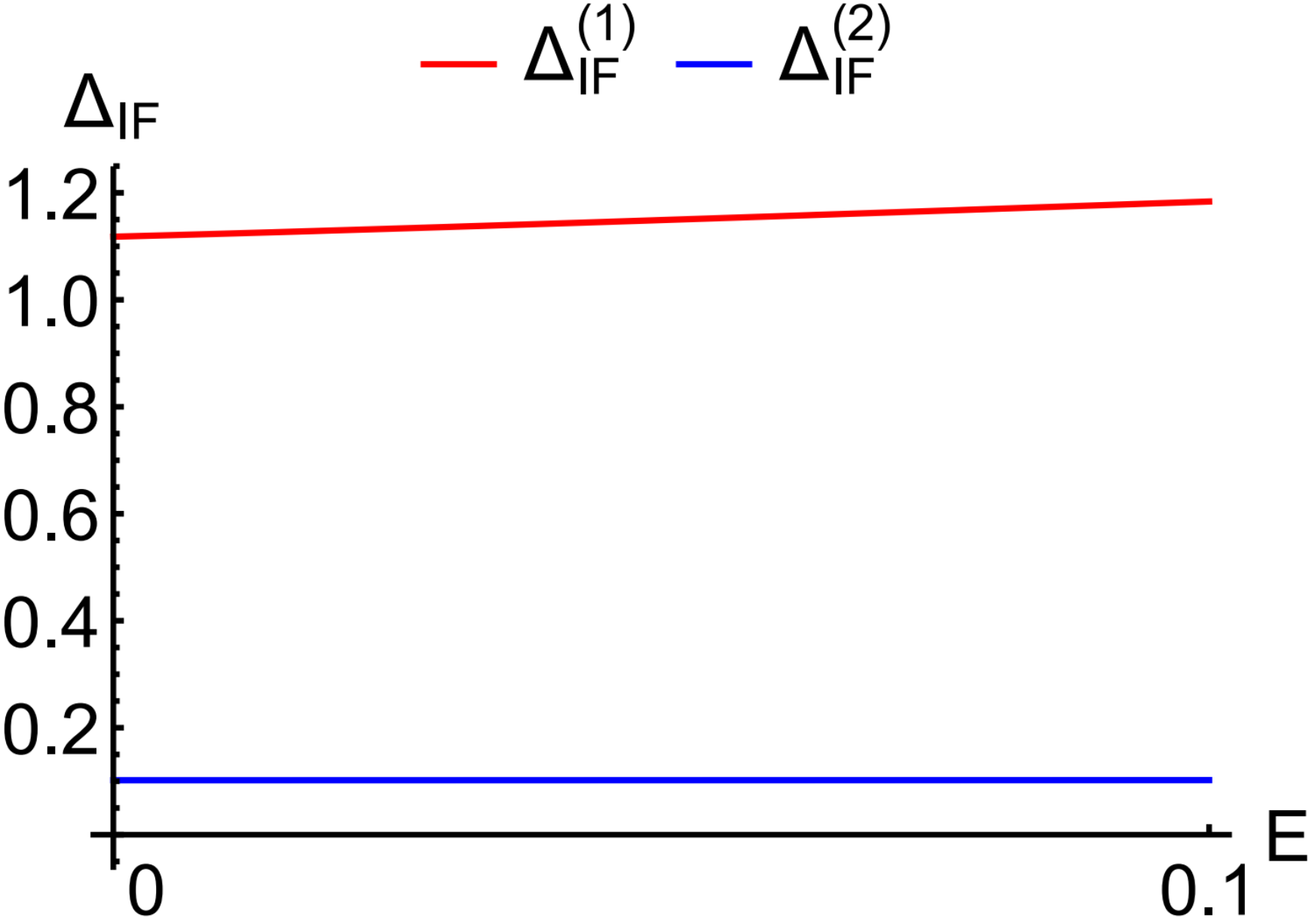}\label{IF_NL5}}
  \caption{(a) The IF shifts $\Delta_{IF}^{(1)}$ and (b) $\Delta_{IF}^{(2)}$ as a function of $\theta$ and $\phi$, with $E=0.01$. (c,d,e) Dependence of the IF shifts on $\theta$, $\phi$ and $E$, respectively, with the other two parameters fixed. The fixed parameters are: $\phi=0.2$ and $E=0.1$ in (c), $\theta=0.5$ and $E=0.01$ in (d), and $\theta=0.1$ and $\phi=0.2$ in (e).} \label{Fig_IFNL}
\end{figure}

The IF shifts are calculated numerically and shown in Fig.\ref{Fig_IFNL}. The vanishing of both $\Delta_{IF}^{(1)}$ and $\Delta_{IF}^{(2)}$ at $\theta=\pi/2$ is due to the vanishing of the Berry curvature in the $k_z=0$ plane, while the vanishing of $\Delta_{IF}^{(2)}$ at $\theta=0$ is due to the equality of $k_x^I$ and $k_x'$. The divergence at $\phi\to\pi/2$ is explained by the singularity of the Berry curvature that is not cancelled out by the smallness of the integral range. Note that both $\Delta_{IF}^{(1)}$ and $\Delta_{IF}^{(2)}$ have a weak dependence on $E$, in contrast to the IF shift in pseudospin-$N/2$ semimetals which has a $1/E$ dependence. The reason can be seen from the approximate form of the Berry curvatures Eq.\ref{eq:BCx}--\ref{eq:BCz} at low energy, $\Omega_x\approx-\frac{\sin{2\theta}\cos{\phi}}{4m_z E_+}$, $\Omega_y\approx-\frac{\sin{2\theta}\sin{\phi}}{4m_z E_+}$, and $\Omega_z\approx-\frac{\cos^2{\theta}}{2m_r E_+}$, which are proportional to $1/E$, in contrast to the more singular behavior of the Berry curvatures in pseudospin-$N/2$ semimetals which are proportional to $1/E^2$.

\subsection{Detecting topological Lifshitz transitions by the IF shift}
The vortex ring model Eq.\ref{eq:Hvr} undergoes a topological phase transition from NLSM to WSM phase when $\al\equiv k_0^2$ is swept from $\al>0$ to $\al<0$. In the latter phase, a pair of Weyl points with opposite chirality appear at $(0,0,\pm\sqrt{|\al|})$, around which the low energy Hamiltonian is
\begin{eqnarray}
  H_{VR}\approx \mp\sqrt{|\al|}(\frac{1}{m_z}k_x\si_x+\frac{1}{m_z}k_y\si_y+\frac{1}{m_r}(k_z\mp\sqrt{|\al|})\si_z).\nonumber\\
\end{eqnarray}
Multiple topological Lifshitz transitions\cite{Volovik2017} during this phase transition occur: the low energy Fermi surface changes its topology from a torus ($\al>0$), via a sphere ($\al=0$), to two spheres ($\al<0$)\cite{Lim2017}.

Now we consider how the IF shift depends on the Fermi surface topology within this model. In the phase with a toroidal Fermi surface, we have shown that the IF shift is zero in case (i), whereas in case (ii) the incident beam splits into two reflected beams with different IF shift. If $\al=0$, the IF shift in case (i) is still zero, and in case (ii) it is finite but the beam does not split. In the WSM phase, in case (i) the beam splits into two beams with different IF shift after the reflection, while in case (ii) it does not split\cite{Wang2017}. Therefore, detecting the IF shift can be used to identify topological Lifshitz transitions.

\section{Experimental realization}\label{SecIV}
\begin{figure}[t]
  \centering
  \includegraphics[width=7cm]{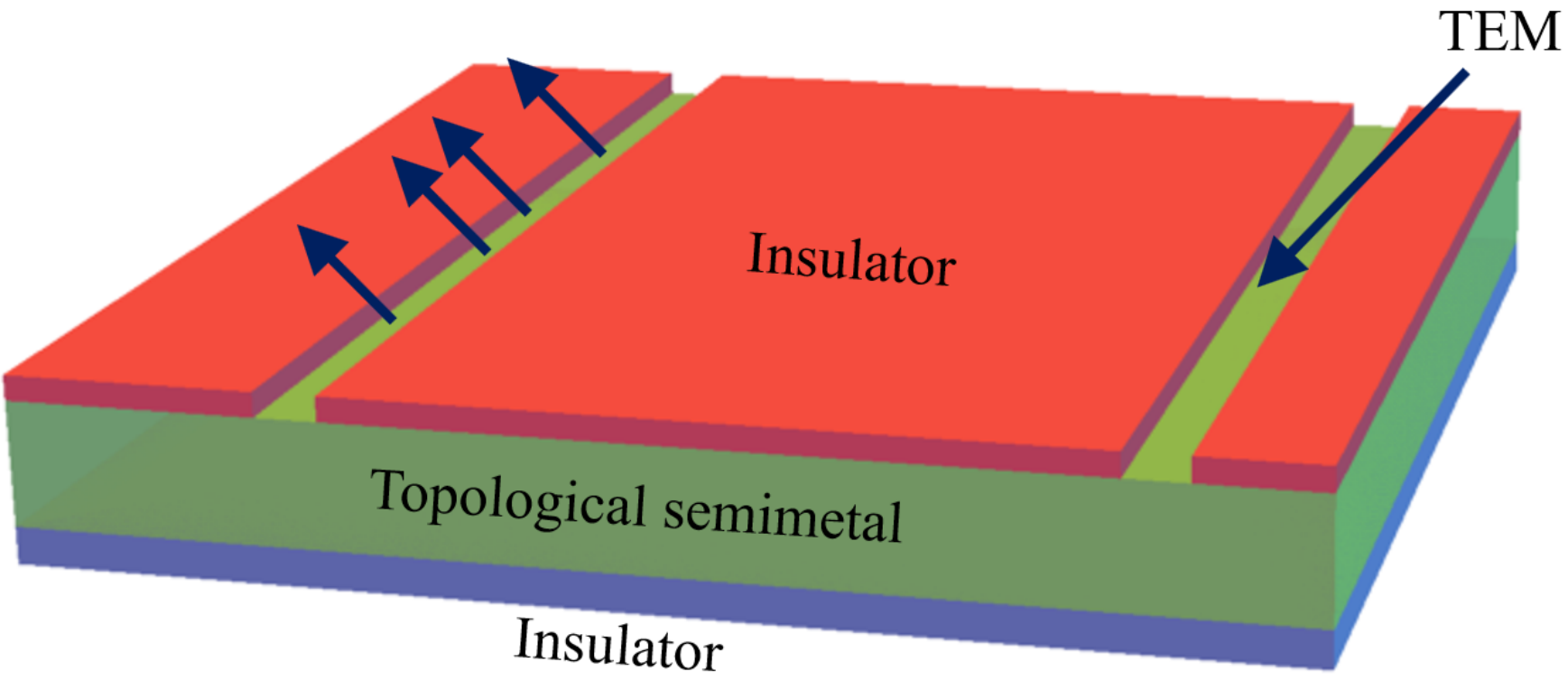}\label{Experiment}
  \caption{Schematic of the experimental design for the detection of the IF shift in a pseudospin-$N/2$ semimetal.}\label{Fig_exp}
\end{figure}
The experimental design for the detection of the IF shift in WSM has been proposed in Ref.\cite{Jiang2015,Yang2015}. More than one WSM with different chemical potential is required in these designs, and WSMs with cylindrical geometry is needed in Ref.\cite{Yang2015}. Here, we propose a simpler design, in which a single sample of a topological semimetal is needed. For the convenience of description, we focus on the detection of the IF shift in pseudospin-$N/2$ semimetals below, but the experimental design also works for NLSMs. As shown in Fig.\ref{Fig_exp}, the semimetal is covered by two different insulators above and below it. The insulating film above has two slits, allowing electrons emitted from the transmission electron microscope (TEM) to go into and out of the semimetal. The energy of the electrons is in the gap of both insulators, thus the electrons are not permitted to transmit in them. After the electron beam goes into the semimetal from one slit, it splits into sub-beams with different pseudospin components. Each sub-beam experiences total reflection every time it hits the interface between the semimetal and the insulating films, during which the IF shift occurs. The IF shift should be smaller than the results we have derived, since the motion of the evanescent wave in the insulating film does not contribute. Note that for a specific sub-beam, the shift at the upper interface $\Delta_{IF}^a$ is in the opposite direction to that at the lower interface $\Delta_{IF}^b$, so the insulators have to be different to leave a net shift $|\Delta_{IF}^a|-|\Delta_{IF}^b|$. After $m$ times of total reflection at the lower interface and $m-1$ times at the upper one, the sub-beam eventually comes out from the other slit, the transverse shift is accumulated to $\Delta_t=m(|\Delta_{IF}^a|-|\Delta_{IF}^b|)-|\Delta_{IF}^a|$. $\Delta_t$ is approximately proportional to $m$ at large $m$. Assume the incident angle is $\theta^I$, the height of the semimetal sample is $h$ and the separation between the two slits is $L$, then $m=L\cot\theta^I/(2h)$. The separation between two outgoing sub-beams is also proportional to $m$. Therefore, by adjusting the geometry of the semimetal sample and the incident angle to make $m$ large, the IF shift should be observable.

As Weyl points have been realized in photonic crystals\cite{Lu2013,Lu2015,Wang2016,Yang2018}, pseudospin-1, pseudospin-3/2 and nodal-line band structures for photons can also be realized. The IF shift in photonic crystals with Weyl points has been investigated in Ref.\cite{Wang2017a}. In Fig.\ref{Fig_exp}, replacing the insulating films with photonic crystals with a band gap, and the semimetal with a gapless topological photonic crystal with pseudospin-$N/2$ or nodal-line band structures, then the IF shift of a beam of light can be observed.

\section{Summary}\label{SecV}
In summary, we have studied the IF shift in pseudospin-$N/2$ semimetals with an arbitrary positive integer $N$ and in NLSMs. The main method we use is the semiclassical equations of motion, while the conservation of the angular momentum assists us to interpret our results in the cases with rotational symmetries. We find that the IF shift of topological pseudospin-$N/2$ fermions depends on the pseudospin components quadratically, with the sign depending on the chirality. By investigating the IF shift in NLSMs with different symmetries, we find it strongly depends on the symmetry of the system. Additionally, we show that the IF shift can be a powerful tool to detect topological Lifshitz transitions. Finally, we propose experimental designs to detect the IF shift in topological semimetals as well as in photonic crystals with topological degeneracies.

\begin{acknowledgements}
This work was supported by NKRDPC-2017YFA0206203, NSFC-11574404, NSFG-2015A030313176, National Supercomputer Center in Guangzhou, Three Big Constructions Supercomputing Application Cultivation Projects and Leading Talent Program of Guangdong Special Projects.
\end{acknowledgements}

\bibliography{IFbib}

\end{document}